\documentstyle[12pt]{article}
\def\dz{\partial_z}
\def\cdz{\partial_{\bar{z}}}
\def\m{\mu_z{}^{\bar{z}}}
\def\cm{\mu_{\bar{z}}{}^z}
\def\cD{\cal D}
\def\e{\vec{e}}
\def\p{\vec{p}}
\def\s{\prime}
\def\tz{\tilde{z}}
\begin{document}
\section*{}
LMU--TPW 94--10 \hfill \\
July 1994 \hfill \\[3ex]
\begin{center}
\Large\bf
Non--Standard Fermion Propagators from Conformal Field Theory
\end{center}
\vspace{-6ex}
\section*{}
\normalsize \rm
\begin{center}
{\bf Rainer Dick}\\ {\small \it Sektion Physik der
Universit\"at M\"unchen\footnote{Address after 27 September 1994:
School of Natural Sciences, Institute
for Advanced Study, Olden Lane, Princeton, NJ 08540, USA.
} \\
Theresienstrasse 37, 80333 M\"unchen,
Germany}
\end{center}
\vspace{-6ex}
\section*{}
{\bf Abstract}:
It is shown that Weyl spinors in 4D Minkowski space are composed of
primary fields of half--integer conformal weights. This
yields representations of fermionic 2--point functions in terms of
correlators of primary fields with a factorized transformation
behavior under the Lorentz group. I employ this observation to
determine the general structure of the corresponding Lorentz
covariant correlators by methods similar to the methods employed
in conformal field theory to determine 2-- and 3--point functions
of primary fields. In particular, the chiral symmetry breaking terms
resemble fermionic 2--point functions of 2D CFT up to a function
of the product of momenta.

The construction also permits for the formulation of covariant meromorphy
con\-straints on spinors in 3+1 dimensions.
\section*{}\begin{center}
\small\it to appear in the proceedings of the Feza G\"ursey
Memorial Conference $I$:\\ On Strings and Symmetries\end{center}
\newpage
\section{Introduction}
The quest for a four--dimensional notion of analyticity and the related
problem to define four--dimensional analogues of two--dimensional conformal
field theories is a subject of much interest and under intense study
since many years.
Already in 1956
Feza G\"ursey employed the quaternionic
formulation of the Dirac equation to derive a conformally invariant
nonlinear spinor equation
\cite{gursey}. He later returned
to the topic of quaternionic
analyticity several times yielding important insights and results
\cite{gt,gj}. Finally, his life--long
fascination in this topic and the harmonic space approach developed
by the Dubna group \cite{gikos,gios}
merged in a remarkable and beautiful
recent paper with Mark Evans and Victor Ogievetsky \cite{ego},
see also \cite{victor}. In this paper the relevance and applicability
of quaternionic analyticity in the framework of self--dual theories
is worked out very clearly.

Notions of quaternionic analyticity arise in a similar manner also in the
twistor approach
to space--time \cite{pr}.

Other Ans\"atze, which were developed recently, employ
direct transfers of methods and notions of 2D conformal field theory
to 4D conformal field theories. This
led e.g.\
to the discovery of
structures reminiscent
of Zamolodchikov's $c$--theorem \cite{c1,jo}, and to new results
on correlation functions in 4D conformal field theories,
including in particular an extension of the central
charge of 2D conformal field theory to a triple of central
charges in 4 dimensions \cite{op,o2}.
Of related interest is the impressive list of
recent results on quasi--primary
fields in the $O(N)$ $\sigma$--model for $2 < d < 4$ \cite{ruehl}.

Here I would like to introduce still another approach to
analyticity in 3+1 dimensions:
I would like to point out that
left or right handed massless spinors in 3+1
dimensions can be interpreted as half--differentials on spheres in
{\sl momentum space}. This implies the possibility to formulate
covariant phase space constraints on spinors of definite helicity
in terms of (anti--)meromorphy constraints.
More specifically, the entries of a spinor
of negative helicity are shown to yield
local representations $\psi(z,\bar{z},E)$
of a primary field of weight $(\frac{1}{2},0)$,
where $z(p)$ denotes stereographic coordinates in momentum space:
\[
z = \frac{p_+^{}}{|\vec{p}\,|-p_3}
\]

The Weyl equation then appears
as a particular consequence
of the transformation behavior under holomorphic
reparametrizations:
\begin{equation} \label{tra}
\psi^{\prime}(z^{\prime},\bar{z}^{\prime},E^{\prime})
=\psi(z,\bar{z},E)
\left(\frac{\partial z^{\prime}}{\partial z}\right)^{-\frac{1}{2}}
\end{equation}
Covariance of the construction follows because
Lorentz transformations induce via $SL(2,\mbox{\bf C})$
holomorphic transformations
of spheres in momentum space, and the resulting transformation behavior
of left handed spinors complies with the corresponding transformations
of half--differentials.
The construction implies in particular, that
left handed spinors can be subjected to covariant constraints
\begin{equation} \label{hol}
\frac{\partial \psi}{\partial\bar{z}}=0
\end{equation}
stating that a left handed spinor which does not depend on $\bar{z}(p)$
in one inertial frame will also remain independent
of this particular combination
of momenta in any other inertial frame.

Another source of motivation for the present work
besides the formulation of covariant analyticity constraints is due
to applications to low energy QCD:

The relevance of methods of 2D field theory in certain kinematical
regimes or large $N$ expansions of QCD
has been noticed in many places, and
applications of notions or techniques of 2D field theory
have proven fruitful recently. In particular,
constructions of
effective 2D field theories to describe high energy
scattering in QCD were given in \cite{lip,ver}.

In the present context,
the expansion of massless
spinors in terms of half--differen\-tials may
provide new insights into the issue
of chiral symmetry breaking in low energy QCD.
For an explanation of this note that
the isomorphy between chiral Weyl spinors on the one hand and
half--differentials on the other hand offers the possibility to write
correlators of massless fermions as a sum of
correlators of primary fields with a factorized transformation behavior
under Lorentz transformations.
If one employs the hypothesis, that any massless fermion propagator has
a representation in terms of correlators of Weyl spinors, as specified
in Eq.\ (\ref{offprop}) below, then
this offers a possibility to apply techniques
of 2D conformal field theory to determine the structure of the propagators
from Lorentz covariance. It turns out that the general Lorentz covariant
propagator in the massless limit is
determined up to 2 functions $f_1^{}$ and
$f_2^{}$ which depend on single, but different arguments \cite{dick1}:
\begin{equation}\label{result}
\langle\Psi(\p\,)\overline{\Psi}(\p^{\,\prime})\rangle=
\end{equation}
\[
\left(\begin{array}{cc}0&1\\0&0\end{array}\right)\otimes
\left(\begin{array}{cc}\bar{z}z^\prime & \bar{z}\\
z^\prime&1\end{array}\right)\langle\phi(\p\,)\phi^+(\p^{\,\prime})\rangle+
\left(\begin{array}{cc}0&0\\1&0\end{array}\right)\otimes
\left(\begin{array}{cc}1 & -\bar{z}^\prime\\
-z& z\bar{z}^\prime
\end{array}\right)\langle\psi(\p\,)\psi^+(\p^{\,\prime})\rangle
\]
\[
+
\left(\begin{array}{cc}1&0\\0&0\end{array}\right)\otimes
\left(\begin{array}{cc}\bar{z} & -\bar{z}\bar{z}^\prime\\
1&-\bar{z}^\prime
\end{array}\right)\langle\phi(\p\,)\psi^+(\p^{\,\prime})\rangle +
\left(\begin{array}{cc}0&0\\0&1\end{array}\right)\otimes
\left(\begin{array}{cc}z^\prime & 1\\
-zz^\prime&-z\end{array}\right)\langle\psi(\p\,)\phi^+(\p^{\,\prime})\rangle
\]
\begin{equation}\label{f1}
\langle\psi(\p_1^{})\psi^+(\p_2^{})\rangle=
\langle\phi(\p_2^{})\phi^+(\p_1^{})\rangle=
f_1^{}\!\left(\frac{|\p_1^{}|}{|\p_2^{}|}\right)
\frac{1+z_1^{}\bar{z}_2^{}}{\sqrt{|\p_1^{}||\p_2^{}|}}\,
\delta_{z\bar{z}}^{}(z_1^{}-z_2^{})
\end{equation}
\begin{equation}\label{f2}
\langle\psi(\p_1^{})\phi^+(\p_2^{})\rangle=
\overline{\langle\phi(\p_2^{})\psi^+(\p_1^{})\rangle}=
\frac{1}{z_1^{}-z_2^{}}\,
f_2^{}\!\left(|\p_1^{}||\p_2^{}|
\frac{(z_1^{}-z_2^{})(\bar{z}_1^{}-\bar{z}_2^{})}
{(1+z_1^{}\bar{z}_1^{})(1+z_2^{}\bar{z}_2^{})}\right)
\end{equation}

Applications of this result to low energy QCD arise from expectations that
chiral symmetry is broken even in the massless limit of QCD,
since otherwise it would be hard to understand that the chiral condensates
of all the light flavours seem to have the same order of magnitude
\cite{yndurain}.
Now the derivation of Eq.\ (\ref{result}) given
in Sec.\ 4 implies that the massless
limit of the light quark propagators must be of this form
with non--vanishing $f_2^{}$--terms, since the terms
containing $f_2^{}$ are the only terms which comply both with Lorentz
covariance and chiral symmetry breaking. Note the consistency of this
result: Since (\ref{result}) provides the general form of a
Lorentz covariant {\sl massless} propagator, those parts of it which
break chiral symmetry necessarily must also break translational
invariance. This is in agreement with Eq.\ (\ref{f2}), since the right
hand side of this equation cannot accomodate for a $\delta$--function
in external momenta. The $f_1^{}$--terms in turn preserve chiral symmetry:
They
do not contribute to a chiral condensate and anticommute with $\gamma_5^{}$.
Consistency of the result in this sector is expressed by the fact that these
terms contain
a $\delta$--function which restricts the correlator to parallel momenta.

Thus our result lends
some support to conjectures that chiral symmetry breaking may appear
as a consequence of confinement:
Accompanying valence quarks yield
background gauge fields from the point of view of the quark described by the
propagator (\ref{result}), and terms which result from breaking of
translational invariance also break chiral symmetry.

Remarkable progress in the study
of confinement has been achieved recently
due to the work of
Seiberg
and Witten, who provide strong evidence for monopole condensation
if $N=2$ supersymmetric Yang--Mills theory is broken down
to $N=1$ supersymmetry
\cite{sw}.

The notion of primary fields will be introduced in a fully covariant
setting in Sec.\ 2, while the isomorphy between Weyl spinors of definite
helicity and half--differentials will be the topic of Sec\ 3.
Sec.\ 4 essentially comprises a group theoretical investigation of
propagators of massless fermions by means of primary field techniques
to yield the result (\ref{result}).
Sections 3 and 4 can be read independently of Sec.\ 2, if the reader
is familiar with the notion of half--differentials
and not interested in a covariant definition of primary
fields. Other readers with primary interest in the
derivation of the massless propagator may be willing to take
Eq.\ (\ref{tra}) as a definition and refer to Sec.\ 2 only if necessary.

\section{Covariant Primary Fields}
In two--dimensional field theories two apparently different formulations
of covariance existed in parallel for several years. On the one hand
two--dimensional field theories can be formulated covariantly in the
usual way employing tensor and spinor fields,
while on the other hand
it is known that in a conformal gauge primary fields can be employed
to ensure covariance with respect to the conformal remnant of the
diffeomorphism group \cite{bpz}.
This was puzzling, because there exist primary fields of half--integral
order on two--manifolds, and it was not clear in
what sense these could be considered as remnants
of tensor or spinor fields in a conformal
gauge\footnote{In this section, spinor refers to 2D spinors}.
Furthermore, it
was unclear how half--differentials should transform under non--conformal
transformations, or how they could be defined outside the realm of
conformal gauge fixing.
The puzzle was partially solved by the introduction of a
covariant definition of primary fields \cite{dick},
thus demonstrating
that primary fields yield factorized
representations of the full two--dimensional diffeomorphism group.
This work also included a demonstration of
isomorphy between tensor fields and covariant primary fields of integer
weight. However, the exact relation between spinors in two dimensions
and the covariant
half--differentials of \cite{dick} was given only recently in \cite{nic},
where the formalism was further developed and
applied to two--dimensional supergravity.

Initially primary fields $\Phi$ of conformal weight
$(\lambda,\bar{\lambda})$
on a two--manifold ${\cal M}$
are defined by their transformation behavior
under a holomorphic change of charts $z\to u(z)$ \cite{bpz}:
\begin{equation} \label{dpa}
\Phi(u,\bar{u}) = \Phi(z,\bar{z})\cdot \left({\frac{\partial
u}{\partial z}}\right)^{-\lambda} \cdot \left({\frac{\partial \bar{u}}{\partial
\bar{z}}}\right)^{-\bar{\lambda}}
\end{equation}
where I
employed the usual convention to denote the weight for the complex
conjugate sector of coordinates by $\bar{\lambda}$.

The scaling dimension of the field $\Phi$ is
$\Delta = \lambda+ \bar{\lambda}$
and the
spin\footnote{We distinguish between the spin $\sigma$ referring
to rotations induced by diffeomorphisms of ${\cal M}$ and the spin
$s$ referring to rotations of tangent frames.}
is $\sigma = \lambda - \bar{\lambda}$.
A cohomological investigation reveals that
$\sigma$ is restricted
to integer
or half--integer values, while no similar restriction is imposed on
the scaling dimension. We will demonstrate this in the more general setting
of covariant primary fields below.

The factorized transformation behavior makes primary fields particular
convenient for the formulation of two--dimensional field theories and
the investigation of short distance expansions.
However, this definition of primary fields
works only in a conformal gauge, i.e.\ in an atlas
with holomorphic transition functions. This causes no problem for integer
values of $\lambda$ and $\bar{\lambda}$, because the corresponding primary
fields might be considered as remnants of tensor fields in the conformal gauge.
However, such an interpretation is not possible for fractional conformal
weights. Furthermore, if the metric of the two--manifold $\cal M$ is
considered as a dynamical degree of freedom it is very inconvenient to
switch to a conformal gauge, because this implies that two degrees of freedom
of the metric corresponding to the Beltrami--parameters (see below)
are hidden in the holomorphic transition functions.
Therefore, in a conformal gauge
it is impossible to formulate the dynamics
of the metric in terms of local fields.

To avoid the restriction to conformal atlases
requires a generalization of equation (\ref{dpa})
to diffeomorphisms
$z \to u(z,{\bar{z}})$, i.e.\ we will define primary fields for arbitrary
atlases on smooth two--manifolds, thereby introducing a covariant definition
of half--differentials. Hence, in the sequel $z,w$ and $u$ will denote
complex local coordinates, but no holomorphy conditions on transformations
will be assumed any more.
To define covariant primary fields
it is convenient to switch to a Beltrami--parametrization
of the metric:
\begin{equation} \label{metric}
(ds)^2={\frac{2\,g_{z\bar{z}}}{1+\mu_{\bar{z}}{}^z\cdot\mu_z{}^{\bar{z}}}}
\cdot \left|dz+\mu_{\bar{z}}{}^z\cdot d\bar{z}\right|^2
\end{equation}
i.e.\ the Beltrami--parameters $\{\m,\cm\}$
specify the metric modulo scaling
transformations:
\begin{eqnarray} \label{bel1}
\cm &=& \frac{g_{z\bar{z}}^{}-
\sqrt{g_{z\bar{z}}^2-g_{zz}^{}g_{\bar{z}\bar{z}}^{}}}
{g_{zz}^{}}\\
{}&=& \frac{g_{\bar{z}\bar{z}}^{}}{g_{z\bar{z}}^{}+
\sqrt{g_{z\bar{z}}^2-g_{zz}^{}g_{\bar{z}\bar{z}}^{}}} = \mu_z{}^{\bar{z}*}
\nonumber\\ \label{bel2}
\frac{g_{zz}^{}}{g_{z\bar{z}}^{}} &=& \frac{2\mu_z{}^{\bar{z}}}
{1+\mu_{\bar{z}}{}^z\mu_z{}^{\bar{z}}}
\end{eqnarray}
The Beltrami--parameters satisfy $\mu_{\bar{z}}{}^z\mu_z{}^{\bar{z}} < 1$
and have a subtle transformation behavior under reparametrizations
$z\to u(z,\bar{z})$ with
$\left|\partial_z u\right| >
\left| \partial_{\bar{z}} u\right| $:
\begin{equation} \label{trb}
\mu_{\bar{u}}{}^u = \frac{\mu_{\bar{z}}{}^z \cdot \partial_z u
- \partial_{\bar{z}} u}{\partial_{\bar{z}} \bar{u} - \mu_{\bar{z}}{}^z \cdot
\partial_z \bar{u}}
= \frac{\partial_{\bar{u}} z + \mu_{\bar{z}}{}^z \cdot \partial_{\bar{u}}
\bar{z}}{\partial_u z + \mu_{\bar{z}}{}^z \cdot \partial_u \bar{z}}
\end{equation}
This transformation law implies in particular
\[
\partial_{\bar{z}}-\cm \dz = (\cdz \bar{u} - \cm\dz \bar{u})
(\partial_{\bar{u}} - \mu_{\bar{u}}{}^u \partial_u) =
\frac{1}{\partial_{\bar{u}}\bar{z} - \mu_{\bar{u}}{}^u \partial_u\bar{z}}
(\partial_{\bar{u}} - \mu_{\bar{u}}{}^u \partial_u)
\]

This observation motivates the introduction of particular
non--holonomic bases of vector fields and differentials on two--manifolds
$\cal M$:
\begin{eqnarray} \label{pbs1}
{\cal D}_z &=& \partial_z - \mu_z{}^{\bar{z}} \cdot
\partial_{\bar{z}} \\*[1ex] \label{pbs2}
{\cal D}z &=& \frac{1}{1-\mu_{\bar{z}}{}^z \cdot \mu_z{}^{\bar{z}}}
\left(dz + \mu_{\bar{z}}{}^z \cdot d{\bar{z}}\right) \\[1ex] \label{pbs3}
\partial_z &=& \frac{1}{1-\mu_{\bar{z}}{}^z\cdot\mu_z{}^{\bar{z}}}
\left({\cal D}_z + \mu_z{}^{\bar{z}} \cdot {\cal D}_{\bar{z}}\right) \\*[1ex]
\label{pbs4}
dz &=& {\cal D}z-\mu_{\bar{z}}{}^z\cdot{\cal D}{\bar{z}}
\end{eqnarray}

These bases are distinguished by their factorized transformation
properties under diffeomorphisms:
\begin{equation} \label{tbs}
{\cal D}_u = \left({\cal D}_u z\right)\,{\cal D}_z \qquad {\cal D}u = {\cal
D}z\,{\cal D}_z u \qquad {\cal D}_z u = \left({\cal D}_u z\right)^{-1}
\end{equation}
thus allowing us to introduce
a consistent covariant definition of primary fields:

A field $\Phi$ over a two--manifold is denoted as {\em primary} of weight
$(\lambda,\bar{\lambda})$
if its local representations $\Phi(z,\bar{z})$
transform under a change of coordinates $z,\bar{z} \to u,\bar{u}$ according to
\[
\Phi(u,\bar u) = \Phi(z,\bar z)\cdot \left({\cal D}_z u\right)^{-\lambda}
\cdot \left({\cal D}_{\bar z} {\bar u}\right)^{-\bar \lambda}
\]

\vspace{2ex}
In particular any tensor representation of the diffeomorphism group factorizes
into appropriate primary fields with integer weights upon expansion with
respect to the non--holonomic bases (\ref{pbs1},\ref{pbs2}),
but the crucial point is that
fractional weights can be defined as well
without conformal gauge fixing.

As we remarked before,
there is a restriction on the admissible values of the weight
$(\lambda,\bar \lambda)$: In a region of three intersecting patches
$U_I^{},U_J^{},U_K{}$ with coordinates
$z_I^{}, z_J^{}, z_K^{}, z_I^{} = f_{IJ}^{}(z_J^{},\bar{z}_J^{})$, etc.,
the product of transition functions for a roundtrip
$z_I^{} \to z_J^{} \to z_K^{} \to z_I^{}$ must yield the identity:
\begin{equation} \label{tcond}
({\cD}_{z_K}f_{IK}^{})^\lambda({\cD}_{z_J}f_{KJ}^{})^\lambda({\cD}_{z_I}
f_{JI}^{})^\lambda
({\cD}_{\bar{z}_K}\bar{f}_{IK}^{})^{\bar\lambda}
({\cD}_{\bar{z}_J}\bar{f}_{KJ}^{})^{\bar\lambda}
({\cD}_{\bar{z}_I}\bar{f}_{JI}^{})^{\bar\lambda} = 1
\end{equation}
For integer weights this condition is automatically fulfilled due to
$f_{KI}^{} = f_{KJ}^{}\circ f_{JI}^{}$ and (\ref{tbs}).
However, if $\Delta=\frac{r}{s},\sigma=\frac{p}{q}$
are representations of $\Delta$ and $\sigma$ in terms
of integers without common divisors,
and if $q\neq 1$,
then
it is a non--trivial problem to fix the $q$--fold ambiguity in the
definition of the transition functions
$
\left({\cal D}_{z_I}f_{JI}^{}\right)^{\lambda}
\cdot \left({\cal D}_{{\bar z}_{I}} {\bar f}_{JI}^{}\right)^{\bar \lambda}
$
in the intersections of all patches in such a manner that the condition
(\ref{tcond}) is fulfilled. To elaborate this further, we split the
transition functions into modulus and phase according to
\[
{\cal D}_{z_J}f_{IJ}^{} = R_{IJ}^{}\exp (i\phi_{IJ}^{})
\]
If we now stick to the convention to choose $R_{IJ}^{\frac{1}{s}}$
positive real
in any intersection $U_I^{} \cap U_J^{}$, then (\ref{tcond}) reduces to
\begin{equation} \label{scond}
\exp (i\sigma\phi_{IK}^{})\cdot \exp (i\sigma\phi_{KJ}^{})\cdot
\exp (i\sigma\phi_{JI}^{}) = 1
\end{equation}
and this defines the choice of phases as a sheaf--cohomological problem:\\
To clarify this define
\begin{equation} \label{ds}
S_{IJK}^{} \equiv
\exp (i\sigma\phi_{IK}^{})\cdot \exp (i\sigma\phi_{KJ}^{})\cdot
\exp (i\sigma\phi_{JI}^{})
\end{equation}
which is an element of $Z_q$.
Consider the sheaf ${\cal M}\times Z_q$ with base manifold $\cal M$
and stalk $Z_q$. An $n$--cochain is a completely antisymmetric
functional of intersections of $n+1$ patches with values
in $Z_q$:
\begin{eqnarray*}
c(U_{I(0)}^{}\cap U_{I(1)}^{}\cap\ldots\cap U_{I(n)}^{})
&=&c_{I(0)I(1)\ldots I(n)}^{}=c_{I(1)I(0)\ldots I(n)}^{-1}\in Z_q
\\
c(\emptyset)&=&1
\end{eqnarray*}
Then there are coboundary operators $\delta_n$ in the
pre--sheaf related to the cover $\{U_I^{}\}$ mapping $n$--cochains
to $(n+1)$--cochains:\\
\begin{eqnarray*}
(\delta_0^{} c)_{IJ}^{} &=& \frac{c_I^{}}{c_J^{}} \\
(\delta_1^{} c)_{IJK}^{} &=& c_{IJ}^{}\frac{1}{c_{IK}^{}}c_{JK}^{} \\
(\delta_2^{} c)_{IJKL}^{} &=& c_{IJK}^{}\frac{1}{c_{IJL}^{}}c_{IKL}^{}
\frac{1}{c_{JKL}^{}}
\end{eqnarray*}
and we have
\[ \delta_{n+1}^{}\delta_n^{}=1\]
Then $S$ as defined in (\ref{ds}) is a closed 2--cochain:
$\delta_2^{}S = 1$.
Unfortunately this does not imply exactness of $S$, because
the phase factors $\exp(i\sigma\phi_{IJ})$ generically do not
satisfy $x^q=1$. On the other hand exactness is what we are seeking,
because in this case
we would have
\begin{eqnarray*}
S_{IJK}^{} &\equiv&
\exp (i\sigma\phi_{IK}^{})\cdot \exp (i\sigma\phi_{KJ}^{})\cdot
\exp (i\sigma\phi_{JI}^{})\\
{}&=& (\delta_1^{} \theta)_{IJK}^{} = \theta_{IJ}^{}\theta_{JK}^{}
\theta_{KI}^{}
\end{eqnarray*}
for some 1--cochain $\theta$ in ${\cal M}\times Z_q$
and we could rescale the phase factors
$\exp (i\sigma\phi_{IJ}^{}) \to \exp (i\sigma\phi_{IJ}^{})\theta_{JI}^{}$
such that the condition (\ref{scond}) could be fulfilled.
Therefore, we may admit only those values for the
denominator $q$ of the spin, which correspond
to a trivial cohomology group $H^2({\cal M},Z_q)$.
However, it is a classical result on two--manifolds that this
cohomology group equals $\emptyset$ for every ${\cal M}$ if and only if
$q=1$ or $q=2$ \cite{hs}.
Hence,
the spin of primary fields over two--manifolds
is restricted to integral or half--integral values.
This implies in particular that the fractional values of conformal
weights
appearing in the conformal grids of minimal models must be combined
into the weights $(\lambda , \bar{\lambda })$
of primary fields such that $\sigma$
is half--integer or integer.
This
rule seems also justified empirically, because it is
in agreement with
the weights appearing in explicit realizations of
minimal models.

Let us now take a closer look at the correspondence between spinors
on the one hand and primary fields
of half--integer weights on the other hand:

As remarked before, the isomorphy between tensors and primary fields
of integer weights is given by expansion with respect to the
anholonomic basis (\ref{pbs1},\ref{pbs2}) \cite{dick}.

The relation between two--dimensional spinors
and covariant half--differentials
has been clarified by employing an appropriate zweibein
formalism \cite{nic}.
Therefore, consider complex orthogonal bases in the tangent
frames:
\[
\vec{e}_{\zeta}^{}=\frac{1}{2}({\e}_1^{}-i{\e}_2^{})
\]
\[
\eta_{\zeta\zeta}^{}=0,\qquad \eta_{\zeta\bar{\zeta}}^{} = \frac{1}{2}
\]
We stick to the convention that greek indices transform under the
symmetry group of the tangent bundle, while latin indices refer to
transformations under diffeomorphisms.
Remember that in the complex orthogonal
bases rotations in the tangent bundle are diagonal:
\[
\Lambda (\phi) = \left(\begin{array}{cc} e^{i\phi} & 0\\
0 & e^{-i\phi}\end{array}
\right)
\]
For spinors we choose a Weyl basis
$\gamma_1^{}=\sigma_1^{},\gamma_2^{}=\sigma_2^{}$
such that the spinor representation of $SO(2)$ is diagonal as well:
\[
S(\phi) = \left(\begin{array}{cc} \exp(\frac{i}{2}\phi) & 0\\
0 & \exp(-\frac{i}{2}\phi)\end{array}
\right)
\]
In the zweibein formalism the Beltrami parameters appear as ratios
of zweibein components: Insertion of
\[ g_{zz}^{}= e_{z}{}^{\zeta}e_{z}{}^{\bar{\zeta}} \qquad
g_{z\bar{z}}^{}=
\frac{1}{2}(e_{z}{}^{\zeta}e_{\bar{z}}{}^{\bar{\zeta}}+
e_{z}{}^{\bar{\zeta}}e_{\bar{z}}{}^{\zeta})
\]
into
(\ref{bel1}) yields
\begin{equation}\label{zb}
e_{\bar{z}}{}^{\zeta} = \cm e_{z}{}^{\zeta}
\end{equation}
Therefore, the primary zweibein which transforms like a primary field
of weight (1,0) under diffeomorphisms is
\[
\varepsilon_z{}^{\zeta} = e_{z}{}^{\zeta} (1-\m\cm )
\]
Equation (\ref{zb}) implies for the inverse zweibein
\begin{equation}\label{izb}
e^{\bar{z}}{}_{\zeta}= -\m e^{z}{}_{\zeta}
\end{equation}
and therefore {\em the diagonal components of the inverse zweibein are
primary fields of weight} $(-1,0)$ {\em and} $(0,-1)$ {\em respectively}:
\[
\varepsilon^z{}_{\zeta} = e^{z}{}_{\zeta} = \frac{1}{\varepsilon_z{}^{\zeta}}
\]
Thus $e^{z}{}_{\zeta}$ transforms under factorized representations
both under the diffeomorphism group and the tangent space rotations.
Therefore the transformation behavior of fractional powers of
$e^{z}{}_{\zeta}$ is well behaved.
More specifically, $(e^{z}{}_{\zeta})^{-\lambda}
(e^{\bar{z}}{}_{\bar{\zeta}})^{-\bar{\lambda}}$ is a primary field
of weight $(\lambda ,\bar{\lambda})$ under diffeomorphisms and a field of
spin
$s = \bar{\lambda}-\lambda$ under tangent space rotations,
and we know by
(\ref{scond}) that $s$ is restricted to integer and half--integer
values.
In particular, the sought for isomorphy between covariant half--differentials
$\psi_{\sqrt{z}}$ of weight $(\frac{1}{2},0)$ and chiral Weyl spinors
$\psi_{\sqrt{\zeta}}$ is \cite{nic}
\begin{equation}\label{iso}
\psi_{\sqrt{z}}^{}\sqrt{e^{z}{}_{\zeta}} = \psi_{\sqrt{\zeta}}^{}
\end{equation}

Having established equivalence between tensors and spinors on the one
hand and covariant primary fields on the other hand, it is also desirable
to develop a covariant primary differential calculus.
Therefore, we introduce a covariant primary derivative $D_z$ which maps
primary fields of weight $(\lambda ,\bar\lambda )$ and spin $s$ into
primary fields of the same spin and weight $(\lambda +1,\bar\lambda )$:
\begin{equation}\label{covder}
D_z^{} \Phi = {\cD}_z^{} \Phi -\lambda \Gamma^z{}_{zz}\Phi
-\bar{\lambda}\Gamma^{\bar{z}}{}_{\bar{z}z}\Phi -is\Omega_z^{}
\Phi
\end{equation}
Covariance of this construction with respect to diffeomorphisms $z \to u(z,
\bar{z})$ and rotations ${\e}_{\zeta}\to {\e}_{\zeta}\exp(-i\phi)$ implies
\begin{eqnarray}\label{conn1}
\Gamma^u{}_{uu}&=&({\cD}_z^{}u)^{-1}\Gamma^z{}_{zz}-({\cD}_z^{}u)^{-2}
{\cD}_z^{}{\cD}_z^{}u\\ \label{conn2}
\Gamma^{\bar{u}}{}_{\bar{u}u}&=&
({\cD}_z^{}u)^{-1}\Gamma^{\bar{z}}{}_{\bar{z}z}-({\cD}_z^{}u)^{-1}
({\cD}_{\bar{z}}^{}\bar{u})^{-1}{\cD}_z^{}{\cD}_{\bar{z}}^{}\bar{u}\\
\label{conn3}
\Omega_u^{} &=&
({\cD}_z^{}u)^{-1}(\Omega_z^{}+{\cD}_z^{}\phi)
\end{eqnarray}
In applications of this formalism in two--dimensional field theory
there frequently appear the anholonomy coefficients of the primary
bases (\ref{pbs1},\ref{pbs2}), because these coefficients
automatically appear as connection coefficients, if conformally gauge fixed
actions like the Ising model or the bosonic string are covariantized
in this formalism \cite{dick}:
\begin{eqnarray*}
[{\cD}_z^{},{\cD}_{\bar z}^{}] &=& C^{\bar{z}}{}_{\bar{z}z}{\cD}_{\bar z}
- C^z{}_{z\bar{z}}{\cD}_z^{}\\
d{\cD}z &=& C^z{}_{z\bar{z}}{\cD}z\wedge {\cD}\bar{z}\\
C^{\bar{z}}{}_{\bar{z}z} &=& \frac{1}{1-\cm\m }({\cD}_{\bar z}^{}\m
-\m {\cD}_z^{}\cm )
\end{eqnarray*}
The commutator of the covariant primary derivatives is then
\begin{equation}\label{comm}
[D_z^{},D_{\bar{z}}^{}]\Phi = (C^{\bar{z}}{}_{\bar{z}z} -
\Gamma^{\bar{z}}{}_{\bar{z}z})D_{\bar{z}}^{}\Phi
-
(C^z{}_{z\bar{z}} -
\Gamma^z{}_{z\bar{z}})D_z^{}\Phi - \lambda{\cal R}_{z\bar{z}}^{}\Phi
+\bar{\lambda}{\cal R}_{\bar{z}z}^{}\Phi - is{\cal F}_{z\bar{z}}^{}\Phi
\end{equation}
with curvature and field strength
\begin{eqnarray*}
{\cal R}_{z\bar{z}}^{} &=& {\cD}_z^{}\Gamma^z{}_{z\bar{z}}
- {\cD}_{\bar z}^{}\Gamma^z{}_{zz} - C^{\bar z}{}_{\bar{z}z}
\Gamma^z{}_{z\bar{z}} + C^z{}_{z\bar{z}}\Gamma^z{}_{zz}\\
{\cal F}_{z\bar{z}}^{} &=& {\cD}_z^{}\Omega_{\bar{z}}^{}-{\cD}_{\bar{z}}
\Omega_z^{}-C^{\bar z}{}_{\bar{z}z}\Omega_{\bar{z}}^{}+
C^z_{z\bar{z}}\Omega_z^{}
\end{eqnarray*}
Thus curvatures consist of
primary fields
of weight (1,1) in this formalism. However, due to the absence of
second order terms in the connection coefficients, $\cal R$ is not
a mere translation of the ordinary curvature tensor into the primary basis.

Similar to the tensor formalism one may impose constraints on the
connection:
The requirement of invariance of the metric under parallel translations
implies
\begin{equation}\label{metcon}
\Gamma^z{}_{zz} = {\cD}_z^{}\ln({\cal G}_{z\bar{z}}^{})-
\Gamma^{\bar z}{}_{\bar{z}z}
\end{equation}
while the requirement of vanishing torsion implies
\begin{equation}\label{torcon}
\Gamma^{\bar z}{}_{\bar{z}z} = C^{\bar z}{}_{\bar{z}z}
\end{equation}
The consistency of the torsion constraint
with (\ref{conn2}) follows easily from the transformation
behavior of the Lie bracket.

On the other hand, one may also impose a zweibein postulate:
\begin{eqnarray*}
D_z^{}e^z{}_{\zeta} &=& 0\\
D_z^{}e^{\bar z}{}_{\bar{\zeta}} &=& 0
\end{eqnarray*}
implying
\begin{eqnarray*}
i\Omega_z^{} &=& \Gamma^z{}_{zz} + {\cD}_z^{}\ln (e^z{}_{\zeta})\\
{}&=& - \Gamma^{\bar z}{}_{\bar{z}z} - {\cD}_z^{}
\ln (e^{\bar z}{}_{\bar{\zeta}})
\end{eqnarray*}
The zweibein postulate implies in particular invariance of the
metric under parallel translations (\ref{metcon}).

In the sequel we will concentrate on primary fields of weights
$(\frac{1}{2},0)$ or $(0,\frac{1}{2})$.
In 2D field theory these half--differentials appear as
fermionic degrees of freedom of superstrings or in the fermionic formulation
of the Ising model. However, as we will see shortly, they were immanent in
the physics literature since a long time in a different guise.

\section{Massless Fermions and Half--Differentials}
For convenience, I employ the Weyl representation for Dirac matrices.
To clarify the relation between spinors in 3+1 dimensions
and half--differentials,
we introduce stereographic coordinates in momentum space:
\begin{eqnarray}\label{defz}
z &=& \frac{p_+^{}}{|\vec{p}\,|-p_3}\\
\tilde{z} &=& -\frac{1}{z} \nonumber
\end{eqnarray}
According to (\ref{tra})
the relation between local representations $\psi(z,\bar{z},|\p\,|)$ and
$\psi(\tilde{z},\bar{\tilde{z}},|\p\,|)$ of a half--differential
of weight $(\frac{1}{2},0)$ is
\begin{equation}\label{weyl}
\psi(\tilde{z},\bar{\tilde{z}},|\p\,|)= -z\psi(z,\bar{z},|\p\,|)
\end{equation}
However, insertion of (\ref{defz}) demonstrates that this
is the Weyl equation for a massless spinor with opposite signs of
chirality and energy:
\begin{equation}\label{antiweyl1}
(|\p\,|+\p\cdot\vec{\sigma})
\left(\begin{array}{c}\psi(z,\bar{z},|\p\,|)\\
\psi(\tilde{z},\bar{\tilde{z}},|\p\,|)\end{array}\right)=0
\end{equation}
Similarly, the relation between local representations of a primary field
of weight $(0,\frac{1}{2})$
\begin{equation}\label{antiweyl}
\phi(\tilde{z},\bar{\tilde{z}},|\p\,|)= \bar{z}\phi(z,\bar{z},|\p\,|)
\end{equation}
is the Weyl equation for a massless spinor of equal signs of energy
and chirality:
\begin{equation}\label{antiweyl2}
(|\p\,|-\p\cdot\vec{\sigma})\left(\begin{array}{c}
\phi(\tilde{z},\bar{\tilde{z}},|\p\,|)\\ \phi(z,\bar{z},|\p\,|)
\end{array}\right)=0
\end{equation}
To complete the proof that local representations of half--differentials
create Weyl spinors as indicated in (\ref{antiweyl1},
\ref{antiweyl2}), it remains to demonstrate that these objects
exhibit a spinorial transformation behavior under the full Lorentz group:
Under parity or time reversal $z(\p\,)$ goes to $-\bar{z}(\p\,)^{-1}$
and thus half--differentials of weight $(\frac{1}{2},0)$ become
half--differentials of weight $(0,\frac{1}{2})$ and vice versa.
Under proper orthochronous Lorentz transformations $\Lambda(\omega)
=\exp(\frac{1}{2}\omega^{\mu\nu}L_{\mu\nu}^{})$,
with $\omega$ the usual set of rotation and boost parameters,
$z(\p\,)$ goes to
\begin{equation}\label{zlor1}
z^{\prime}(\p^{\,\prime})=\overline{U}\circ z(\p\,)=
\frac{\bar{a}z+\bar{b}}{\bar{c}z+\bar{d}}
\end{equation}
if $E=|\p\,|$, and to
\begin{equation}\label{zlor2}
z^{\prime}(\p^{\,\prime})=U^{-1T}\circ z(\p\,)=
\frac{dz-c}{a-bz}
\end{equation}
if $E=-|\p\,|$.\\
Here $U$ is the positive chirality spin representation of $\Lambda$:
\[
U(\omega)=\exp(\frac{1}{2}\omega^{\mu\nu}\sigma_{\mu\nu}^{})
=\left(\begin{array}{cc} a & b\\ c & d\end{array}\right)\in SL(2,C)
\]
The transformation laws (\ref{zlor1},\ref{zlor2}) are
most easily verified through the usual decomposition of $\Lambda$
into rotations and a boost\footnote{The analog of Eq.\ (\ref{zlor1})
in configuration space is known since a long time. However, it seems
to have escaped attention that this implies a covariant notion
of analyticity in momentum space, and that expressing $z$ as a ratio
actually means to introduce half--differentials.}.

Now assume $E=|\p\,|$:
A half--differential $\phi$ of weight $(0,\frac{1}{2})$
then transforms according to (\ref{tra}) into
\[
\phi^\s(z^\s,\bar{z}^\s,|\p^{\,\s}|)=(c\bar{z}+d)\phi(z,\bar{z},|\p\,|)
\]
implying
\[
\left(\begin{array}{c}\phi^\s(\tz^\s,\bar{\tz}^\s,|\p^{\,\s}|)\\
\phi^\s(z^\s,\bar{z}^\s,|\p^{\,\s}|)\end{array}\right)=U\cdot
\left(\begin{array}{c}\phi(\tz,\bar{\tz},|\p\,|)\\
\phi(z,\bar{z},|\p\,|)\end{array}\right)
\]
Thus a half--differential of weight $(0,\frac{1}{2})$ is equivalent to a
spin--$\langle\frac{1}{2},0\rangle$--representation
of the proper orthochronous
Lorentz group $\cal L_+^{\uparrow}$.
Similarly, it is proved that a half--differential of
weight $(\frac{1}{2},0)$ is equivalent to a
spin--$\langle 0,\frac{1}{2}\rangle$--representation
of the proper orthochronous
Lorentz group if $E=|\p\,|$:
\[
\left(\begin{array}{c}\psi^\s(z^\s,\bar{z}^\s,|\p^{\,\s}|)\\
\psi^\s(\tz^\s,\bar{\tz}^\s,|\p^{\,\s}|)\end{array}\right)
=U^{-1\dagger}\cdot
\left(\begin{array}{c}\psi(z,\bar{z},|\p\,|)\\
\psi(\tz,\bar{\tz},|\p\,|)\end{array}\right)
\]

On the other hand, if $E=-|\p\,|$, then this corresponds to
$U\leftrightarrow U^{-1\dagger}$ in the equations above,
and the assignment
of the half--differentials $\phi$ and $\psi$
to representations of $\cal L_+^{\uparrow}$ is changed.

Note that his construction works in both directions:
Half--differentials thus yield Weyl spinors in Minkowski space
and vice versa. In this framework the half--differen\-tials span
projective representations of the Lorentz group, which
are true representations due to the dependence of the half--differentials
on $|\p\,|$.

\section{Propagators}
According to Sec.\ 3 the expansion of a massless spinor in terms
of helicity states can be written
\begin{equation}\label{expand}
\Psi(p)=\left(\begin{array}{c}1\\0
\end{array}\right)
\otimes \left(\begin{array}{c}\bar{z}\\1\end{array}\right)
\phi(z,\bar{z},|\p\,|)+
\left(\begin{array}{c}0\\1
\end{array}\right)
\otimes \left(\begin{array}{c}1\\-z\end{array}\right)\psi(z,\bar{z},|\p\,|)
\end{equation}
which is the usual expansion expressed in terms of half--differentials.
The massless spinor in space--time contains positive and negative frequency
contributions of this kind:
\[
\Psi(x)=\frac{1}{\sqrt{2\pi}^3}\int\frac{d^3\p}{2|\p\,|}\left(\Psi_+^{}(\p\,)
\exp(\mbox{i}p\cdot x)+
\Psi_-^{}(\p\,)
\exp(-\mbox{i}p\cdot x)\right)
\]
Eq.\ (\ref{expand})
yields representations of the corresponding correlation functions
in terms of primary fields:
\begin{equation}\label{prop}
\langle\Psi(\p\,)\overline{\Psi}(\p^{\,\prime})\rangle=
\end{equation}
\[
\left(\begin{array}{cc}0&1\\0&0\end{array}\right)\otimes
\left(\begin{array}{cc}\bar{z}z^\prime & \bar{z}\\
z^\prime&1\end{array}\right)\langle\phi(\p\,)\phi^+(\p^{\,\prime})\rangle+
\left(\begin{array}{cc}0&0\\1&0\end{array}\right)\otimes
\left(\begin{array}{cc}1 & -\bar{z}^\prime\\
-z& z\bar{z}^\prime
\end{array}\right)\langle\psi(\p\,)\psi^+(\p^{\,\prime})\rangle
\]
\[
+
\left(\begin{array}{cc}1&0\\0&0\end{array}\right)\otimes
\left(\begin{array}{cc}\bar{z} & -\bar{z}\bar{z}^\prime\\
1&-\bar{z}^\prime
\end{array}\right)\langle\phi(\p\,)\psi^+(\p^{\,\prime})\rangle +
\left(\begin{array}{cc}0&0\\0&1\end{array}\right)\otimes
\left(\begin{array}{cc}z^\prime & 1\\
-zz^\prime&-z\end{array}\right)\langle\psi(\p\,)\phi^+(\p^{\,\prime})\rangle
\]
Therefore, the 2--point functions on the right hand side
transform under a factorized
representation of the Lorentz group.
This makes this representation very convenient for the investigation
of all correlations $\langle\Psi(p)\overline{\Psi}(p^\prime)\rangle$
which comply with Lorentz covariance.

While the behavior of the parameters $(z,\bar{z},|\p\,|)$ under
rotations is completely specified by (\ref{zlor1}) in the positive
energy case, for boosts we also have to specify the behavior of
$|\p\,|$. For a boost $\exp(-uL_{03}^{})$ we have
\begin{eqnarray*}
z^\prime &=& \exp(-u)z\\
|\p\,^\prime|&=&\frac{|\p\,|}{z\bar{z}+1}\left(\exp(-u)z\bar{z}+\exp(u)\right)
\end{eqnarray*}

We may now determine the 3--dimensional correlators of primary fields appearing
on the right hand side of (\ref{prop}) by methods similar to the
methods employed to fix 2-- and 3--point functions of primary fields
in 2D conformal field theory: Choose a generating set of the symmetry
group, write down the covariance conditions in infinitesimal form
and solve the resulting differential equations.
Lorentz covariance then fixes
the $(\frac{1}{2},0)\otimes (0,\frac{1}{2})$--differential
$\langle\psi(\p_1^{})\psi^+(\p_2^{})\rangle$ up to a function
$f_1^{}(|\p_1^{}|/|\p_2^{}|)$:
\begin{equation}\label{f11}
\langle\psi(\p_1^{})\psi^+(\p_2^{})\rangle=
f_1^{}\!\left(\frac{|\p_1^{}|}{|\p_2^{}|}\right)
\frac{1+z_1^{}\bar{z}_2^{}}{\sqrt{|\p_1^{}||\p_2^{}|}}\,
\delta_{z\bar{z}}^{}(z_1^{}-z_2^{})
\end{equation}
Similarly, the
$(\frac{1}{2},0)\otimes(\frac{1}{2},0)$--differential
$\langle\psi(\p_1^{})\phi^+(\p_2^{})\rangle$ takes the form
\begin{equation}\label{f22}
\langle\psi(\p_1^{})\phi^+(\p_2^{})\rangle=
\frac{1}{z_1^{}-z_2^{}}\,
f_2^{}\!\left(|\p_1^{}||\p_2^{}|
\frac{(z_1^{}-z_2^{})(\bar{z}_1^{}-\bar{z}_2^{})}
{(1+z_1^{}\bar{z}_1^{})(1+z_2^{}\bar{z}_2^{})}\right)
\end{equation}
while invariance under {\bf C}, {\bf P} or {\bf T} implies
\begin{eqnarray*}
\langle\psi(\p_1^{})\phi^+(\p_2^{})\rangle=
\overline{\langle\phi(\p_2^{})\psi^+(\p_1^{})\rangle}\\
\langle\psi(\p_1^{})\psi^+(\p_2^{})\rangle=
\langle\phi(\p_2^{})\phi^+(\p_1^{})\rangle
\end{eqnarray*}
thus establishing the result we were seeking.

As expected, Lorentz symmetry alone complies with a bilocal propagator
in momentum space, and it
restricts the chiral symmetry preserving
part to parallel momenta. On the other hand, chiral symmetry breaking terms
must account for breaking of translational invariance, in agreement
with (\ref{f22}).

The unperturbed result for the on--shell correlation
\[
\langle\psi(\p\,)\overline{\psi}(\p^{\,\prime})\rangle =
-2p\cdot\gamma|\p\,|\delta(\p-\p^{\,\prime})
\]
is recovered from Eqs.\ (\ref{prop}--\ref{f22}) for
$f_1^{}(x)=\delta(x-1)$, $f_2^{}=0$,
and thus asymptotic freedom implies that $f_1^{}(x)$ contains
$\delta(x-1)$ as a summand, while $f_2^{}$ is bound to vanish
for $|\p_1^{}||\p_2^{}|\to\infty$.

Off--shell extensions of (\ref{prop})
can be inferred from the requirement to yield the same
propagator in configuration space:
\begin{eqnarray}
S(x,x^{\prime})&=&\frac{\Theta(t-t^{\prime})}{(2\pi)^3}
\int\frac{d^3\p}{2|\p\,|}
\int\frac{d^3\p^{\,\prime}}{2|\p^{\,\prime}|}\exp(\mbox{i}p\cdot x)
\mbox{i}\langle\Psi(\p)\overline{\Psi}(\p^{\,\prime})\rangle
\exp(-\mbox{i}p^{\prime}\cdot x^{\prime})\nonumber\\
{}&-&
\frac{\Theta(t^{\prime}-t)}{(2\pi)^3}
\int\frac{d^3\p}{2|\p\,|}
\int\frac{d^3\p^{\,\prime}}{2|\p^{\,\prime}|}\exp(-\mbox{i}p\cdot x)
\mbox{i}\langle\Psi(\p)\overline{\Psi}(\p^{\,\prime})\rangle
\exp(\mbox{i}p^{\prime}\cdot x^{\prime})\nonumber\\
{}&=&\frac{1}{(2\pi)^4}\int d^4p
\int d^4p^{\prime}\exp(\mbox{i}p\cdot x)
S(p,p^{\prime})\exp(-\mbox{i}p^{\prime}\cdot x^{\prime}) \label{offprop}
\end{eqnarray}
thus fixing the structure up to the 2 functions $f_1^{}$, $f_2^{}$.
{}From this point of view
chiral symmetry breaking in the massless limit of QCD
amounts to the question
which $f_2^{}$ terms result from confinement,
and how they eventually translate
into massterms of effective theories.
\\[2ex]
{\bf Acknowledgement:}
I would like to thank Hermann Nicolai and Julius Wess
for helpful discussions at various stages of this work.
I would also like to thank the organizers of the conference for
creating a stimulating scientific atmosphere very much in the spirit
of Feza G\"ursey.

\end{document}